%% file: MR_motion_score_v4_ogt.tex
\documentclass[journal]{IEEEtran}
\usepackage{cite}
\usepackage{subfig}
\usepackage{amsmath,amssymb,amsfonts,amsthm}
\usepackage{algorithm}
\usepackage{algpseudocode}
\usepackage{graphicx}
\usepackage{hyperref}
\usepackage{multirow}
\usepackage{multicol}
\usepackage{booktabs}
\usepackage{makecell}
\usepackage{tabularx}
\usepackage{gensymb}
\usepackage{color, colortbl}
\hypersetup{colorlinks=true,urlcolor=red}
\input{macros}

\definecolor{Gray}{gray}{0.9}

\begin{document}

\title{Annealed Score-Based Diffusion Model for MR Motion Artifact Reduction}
\date{\vspace{-4ex}}

\author{Gyutaek Oh, 
		Jeong Eun Lee, 
		and~Jong~Chul~Ye,~\IEEEmembership{Fellow,~IEEE}% <-this % stops a space
\thanks{G. Oh is with the Department of Bio and Brain Engineering, 
		Korea Advanced Institute of Science and Technology (KAIST), 
		Daejeon 34141, Republic of Korea (e-mail: okt0711@kaist.ac.kr).
		J. C. Ye is with the Kim Jaechul Graduate School of AI,
		Korea Advanced Institute of Science and Technology (KAIST), 
		Daejeon 34141, Republic of Korea (e-mail: jong.ye@kaist.ac.kr).
		J.E. Lee is with the Department of Radiology, Chungnam National University Hospital, Chungnam National University College of Medicine, 282 Munhwa-ro, Jung-gu, 
		Daejeon 35015, Republic of Korea (e-mail: leeje290@gmail.com).}
}

\maketitle

% As a general rule, do not put math, special symbols or citations in the abstract or keywords.
\begin{abstract}
Motion artifact reduction is one of the important research topics in MR imaging, as the motion artifact degrades image quality and makes diagnosis difficult.
Recently, many deep learning approaches have been studied for motion artifact reduction.
Unfortunately, most existing models are trained in a supervised manner, requiring paired motion-corrupted and motion-free images, or are based on a strict motion-corruption model, which limits their use for real-world situations.
To address this issue, here we present an annealed score-based diffusion model for MRI motion artifact reduction.
Specifically, we train a score-based model using only motion-free images, and then motion artifacts are removed by applying forward and reverse diffusion processes repeatedly to gradually impose a low-frequency data consistency. % while preserving the high-frequency details.
Experimental results verify that the proposed method successfully reduces both simulated and in vivo motion artifacts, outperforming the state-of-the-art deep learning methods.

\end{abstract}


\IEEEpeerreviewmaketitle

\section{Introduction}\label{sec:introduction}
\IEEEPARstart{M}{agnetic} resonance imaging (MRI) is an imaging technique that provides various types of contrast images without radiation exposure or invasive procedure.
Despite many advantages, MRI requires a long acquisition time due to its imaging physics.
Furthermore, the long acquisition time leads to motion artifacts due to the movement of the patient, so the motion artifact is considered one of the main problems of MRI.

In addition, the contrast agent injection may cause motion artifacts in MRI.
For example, gadoxetic acid (Gd-EOB-DTPA) is one of the liver-specific MRI contrast agents that can help the diagnosis of diseases such as hepatocellular carcinoma, liver metastases \cite{nishie2015detectability,verloh2015liver} by providing hepatobiliary phase (HBP) imaging \cite{kubota2012correlation}.
However, the administration of Gd-EOB-DTPA can occur acute transient dyspnea, resulting in transient severe motion (TSM) \cite{davenport2013comparison}.
If TSM occurs, the image quality of the arterial phase is degraded, and the accuracy of diagnosis can be affected.
So, an algorithm to correct the motion artifact due to TSM is required.

There have been several attempts to reduce the motion artifact of MRI by tracking the motion \cite{white2010promo,todd2015prospective}, or changing sampling trajectory or imaging sequence \cite{liao1997reduction,pipe1999motion,vasanawala2010navigated}.
However, they require additional devices or scan time, and the types of motion artifacts corrected by these methods are limited.

Motion artifact correction algorithms based on compressed sensing (CS) \cite{usman2013motion,yang2013sparse,jin2017mri} also have been investigated.
CS-based algorithms have shown high-quality results, but they have limitations such as the difficulty of hyperparameter tuning and high computational complexity.
Furthermore, many CS algorithms require raw $k$-space data, which are rarely obtained in clinical environments due to storage limitations.

Recent studies for MRI motion artifact reduction are based on deep learning methods \cite{duffy2018retrospective,oksuz2019detection,liu2020motion,tamada2020motion,lyu2021cine,al2022stacked,kuzmina2022autofocusing+}.
Deep learning methods have shown improved performance and reduced run time compared to previous methods.
However, most of the deep learning methods are based on supervised learning approaches.
Since paired motion-free and corrupted data are difficult to obtain, these methods usually utilize simulated motion artifact images to train the networks.
Therefore, it is difficult to apply them to real motion-corrupted data.

To overcome the limitation of simulation-based deep learning methods, deep learning methods using unpaired data have also been explored.
Some methods interpret the motion artifact reduction problem as image-to-image translation \cite{armanious2020unsupervised,liu2021learning}, and address them based on CycleGAN architecture \cite{zhu2017unpaired}.
Although they utilize real motion artifact data, the performance of these algorithms is often limited because there is no explicit motion artifact rejection mechanism.
%Furthermore, the training of architectures similar to CycleGAN is often difficult because they are composed of several networks.

Recently, we proposed an algorithm for MR motion artifact reduction using bootstrap subsampling and aggregation \cite{oh2021unpaired}.
Under the assumption that the motion artifact appears as $k$-space outliers, the method removes the motion artifact by rejecting $k$-space outliers 
in a probabilistic manner.
Although our prior method outperforms other simulation-based or unpaired deep learning methods, there exist limitations if the motion artifact does not appear as sparse outliers in $k$-space.

\begin{figure*}[!t]
\centerline{\includegraphics[width=0.8\textwidth]{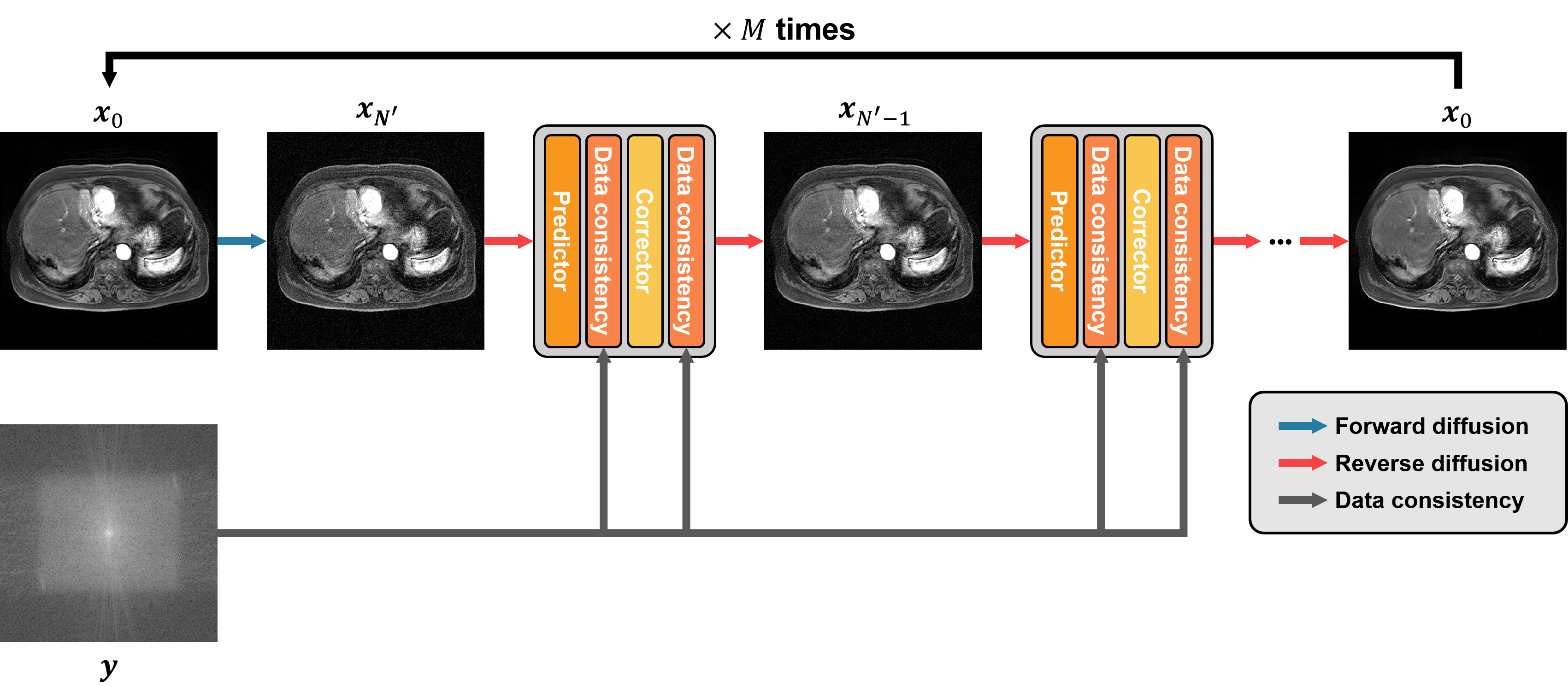}}
\caption{The overall procedure of our method. $\xb_{N'}$ is generated from the motion-corrupted image $\xb_0$ by the forward diffusion.
Then the MR image with reduced motion artifact $\xb_0$ is sampled by solving the reverse-time SDE.
$\yb$ denotes the measurement ($k$-space of motion-corrupted image), and it is used in the data consistency step to prevent the severe deformation of the output image.
The output goes through forward and reverse diffusion iteratively to obtain the final reconstructed image.}
%\vspace{-0.5cm}
\label{fig:sampling}
\end{figure*}

Recently, score-based diffusion models \cite{song2019generative, ho2020denoising, song2020score} have shown remarkable performance in the field of image generation.
In score-based diffusion models, a network that estimates the score, the gradient of the log probability density function, is first trained, and then images can be generated by solving the reverse-time stochastic differential equation (SDE).
Furthermore, it has been verified that unconditionally trained diffusion models can be applied to solve various inverse problems by adjusting sampling procedure using constraints \cite{song2021solving,chung2022come,chung2022improving,chung2022score,saharia2022image}.
Importantly, the unconditionally trained diffusion models do not require paired data, so it is possible to solve inverse problems in an unsupervised manner.

Inspired by this, here we propose a novel MRI motion artifact reduction method using score-based diffusion models.
Fig. \ref{fig:sampling} shows the overall procedure of the proposed method.
%In the proposed method, the score-based diffusion model reconstructs the motion-reduced image through reverse diffusion.
During reverse diffusion, the low-frequency data consistency is gradually imposed
in an iterative manner so that the overall structure of the original image is maintained and helps to remove only motion artifact components.

In particular, our constraint is designed based on the observation that the motion artifacts in MRI usually occurred in the high-frequency region of $k$-space.
This is because $k$-space acquisition is usually performed first in the center region and motion occurs after a certain period after the start of acquisition, so $k$-space samples that include motion artifacts generally appear in high-frequency regions.
Therefore, during the reverse diffusion, the low-frequency region needs to be maintained and only the high-frequency region should be corrected by diffusion sampling.
However, because the high-frequency region of $k$-space also contains the information of details, the detailed structures of images can be altered or vanished if the data consistency step in Eq. \eqref{eq:dc_low} is applied directly.
To address this issue, we propose an annealed reverse sampling procedure where the data consistency step is gradually applied in a repeated manner to maintain high-frequency details of measurements.

%Furthermore, because our score-based diffusion model is trained with only motion-free images, it is a fully unsupervised method.

The remaining parts of the paper are constructed as follows.
Section \ref{sec:backgrounds} introduces backgrounds of score-based diffusion models.
Section \ref{sec:motivation} contains the key idea of the proposed method. 
The experimental setting is explained in Section \ref{sec:methods}, and qualitative and quantitative results are shown in Section \ref{sec:results}.
Section \ref{sec:discussion} and \ref{sec:conclusion} contains the discussion and conclusion of our paper.

\section{Backgrounds}\label{sec:backgrounds}

\subsection{Score-Based Diffusion Models}
A continuous diffusion process can be represented as $\{\xb(t)\}^1_{t = 0}$, where $t\in[0, 1]$ denotes the time variable.
Here, $\xb(0) \sim p_{data}$ where $p_{data}$ is the data distribution, and $\xb(1) \sim p_1$ where $p_1$ refers to the noise distribution which is commonly set to Gaussian distribution.
Then, the diffusion process can be modeled by the solution of the following stochastic differential equation:
\begin{equation}\label{eq:forward}
    d\xb = \fb(\xb, t)dt + g(t)d\wb,
\end{equation}
where $\fb : \Rd^d \rightarrow \Rd^d$ is a drift coefficient, $g : \Rd \rightarrow \Rd$ is a diffusion coefficient, and $\wb$ denotes a standard Wiener process.
By solving Eq. \eqref{eq:forward}, it is possible to transmit a sample from the data distribution to that of the noise distribution through the forward diffusion process.

If it is possible to reverse the diffusion process in Eq. \eqref{eq:forward}, then we can obtain samples of the data distribution from samples of the noise distribution.
In \cite{anderson1982reverse}, it was shown that the reverse process  is also a diffusion process that can be modeled by following reverse SDE:
\begin{equation}\label{eq:reverse}
    d\xb = [\fb(\xb, t) - g(t)^2\nabla_\xb\log{p_t(\xb)}]dt + g(t)d\bar\wb
\end{equation}
where $\bar\wb$ is also a standard Wiener process from time 1 to 0, and $\nabla_\xb\log{p_t(\xb)}$ denotes the score function.
Therefore, if the score function can be estimated, it is possible to derive the reverse diffusion process and generate samples of data distribution from random Gaussian noise.

%It is possible to construct different kinds of SDEs depending on
Among the many possible choices of $\fb$ and $g$, we choose variance exploding SDE (VE-SDE) \cite{song2020score}, where $\fb$ and $g$ are defined by
\begin{equation}
    \fb = 0, \quad g = \sqrt{\frac{d[\sigma^2(t)]}{dt}},
\end{equation}
and
\begin{equation}
    \sigma(t) = \sigma_{\text{min}}\left(\frac{\sigma_{\text{max}}}{\sigma_{\text{min}}}\right)^t.
\end{equation}
Then, the reverse SDE in Eq. \eqref{eq:reverse} can be rewritten as:
\begin{equation}\label{eq:reverse_vesde}
    d\xb = -\frac{d[\sigma^2(t)]}{dt}\nabla_\xb\log{p_t(\xb)}dt + \sqrt{\frac{d[\sigma^2(t)]}{dt}}d\bar\wb.
\end{equation}
Here, the time index $t$ is usually discretized uniformly into $N$ intervals, and $\xb_i$ and $\sigma_i$ can be defined as
\begin{equation}\label{eq:sigma}
    \xb_i := \xb(t)|_{t=\frac{i - 1}{N - 1}}, \quad \sigma_i := \sigma_{\text{min}}\left(\frac{\sigma_{\text{max}}}{\sigma_{\text{min}}}\right)^{\frac{i - 1}{N - 1}}.
\end{equation}

%The score function $\nabla_\xb\log{p_t(\xb)}$ should be estimated to solve the reverse SDE in Eq. \eqref{eq:reverse_vesde}.
The score function $\nabla_\xb\log{p_t(\xb)}$ is generally estimated by training a neural network $\blmath{s_\theta}(\xb(t), t)$ with denoising score matching \cite{vincent2011connection}.
The training of the score-based model with denoising score matching can be done by minimizing the following objective function:
\begin{equation}\label{eq:objective}
    \begin{split}
        \min_{\blmath\theta}\Ed_t\Bigl\{\lambda(t)&\Ed_{\xb(0)}\Ed_{\xb(t)|\xb(0)}\Bigl[\\
        &\norm{\blmath{s_\theta}(\xb(t), t) - \nabla_{\xb(t)}\log{p_{0t}(\xb(t)|\xb(0))}}_2^2\Bigr]\Bigr\}.
    \end{split}
\end{equation}
After training the neural network and plugging it into Eq. \eqref{eq:reverse_vesde}, the reverse SDE can be solved by numerical SDE solvers or predictor-corrector (PC) samplers \cite{song2020score}.

\begin{algorithm}[!t]
\caption{CCDF with PC sampler}\label{al:CCDF}
\begin{algorithmic}[1]
\Require{$\xb_0$, $\yb$, $N'$, $\{\sigma_i\}^{N'}_{i=1}$, $\{\epsilon_i\}^{N'}_{i=1}$, $\blmath{s_\theta}$}
\State{$\zb \sim \Nc(\blmath{0}, \Ib)$}
\State{$\xb_{N'} \leftarrow \xb_0 + \sigma_{N'}\zb$} \Comment{Forward diffusion}
\For{$i = N'$ to 1}
\State{$\xb'_{i - 1} \leftarrow \xb_i + (\sigma^2_i - \sigma^2_{i - 1})\blmath{s_\theta}(\xb_i, \sigma_i)$}
\State{$\zb \sim \Nc(\blmath{0}, \Ib)$}
\State{$\xb_{i - 1} \leftarrow \x'_{i - 1} + \sqrt{\sigma^2_i - \sigma^2_{i - 1}}\zb$} \Comment{Predictor}
\State{$\xb_{i - 1} \leftarrow \text{Data consistency}(\xb_{i - 1}, \yb)$\\\qquad} \Comment{Data consistency}
\State{$\zb \sim \Nc(\blmath{0}, \Ib)$}
\State{$\xb_{i - 1} \leftarrow \xb_{i - 1} + \epsilon_i\blmath{s_\theta}(\xb_i, \sigma_i) + \sqrt{2\epsilon_i}\zb$} \Comment{Corrector}
\State{$\xb_{i - 1} \leftarrow \text{Data consistency}(\xb_{i - 1}, \yb)$\\\qquad} \Comment{Data consistency}
\EndFor
\end{algorithmic}
\end{algorithm}

\subsection{Come-Closer-Diffuse-Faster (CCDF)}
The main drawback of score-based diffusion models is their slow sampling time.
Because the sampling  starts from the random Gaussian noise and usually requires thousands of steps, the sampling time of score-based diffusion models is too long.
In the prior work \cite{chung2022come}, the authors proposed a method called Come-Closer-Diffuse-Faster (CCDF) to reduce the sampling time of diffusion models when solving inverse problems.
Specifically, instead of starting sampling from random Gaussian noise, the forward diffusion is first applied from the initial reconstruction, leading to only few steps of reverse diffusion to get the final reconstruction.

More specifically, Algorithm \ref{al:CCDF} shows the CCDF sampling procedure using the PC sampler, where $\yb$ denotes the initial measurement, and $N' = Nt'$ is the number of reverse diffusion steps where $t' \in [0, 1]$.
Here, the data consistency step should be non-expansive to maintain the stochastic contraction mapping nature of reverse diffusion sampling \cite{chung2022come}.
With a better initialization followed by one-step forward diffusion, CCDF largely reduces the reverse sampling time for solving inverse problems \cite{chung2022come}.

\section{Main Contribution}\label{sec:motivation}
\subsection{Motivation}
In our prior work \cite{oh2021unpaired}, we solved the motion artifact reduction problem by regarding motion artifacts as sparse outliers in $k$-space.
%Because the acquisition time along the frequency encoding direction is relatively short, it can be assumed that the motion does not occur during the frequency encoding.
Specifically, if the motion is occurred by translation or rotation, it is assumed to result in $k$-space phase shift or rotation at the specific phase encoding lines:
%Then, the motion-corrupted $k$-space can be represented as follows:
\begin{equation}\label{eq:motion}
    \hat\yb(k_x, k_y) = 
    \begin{cases}
    \Fb\{\Rb_\alpha{\xb}\}e^{-j\Phi}, & k_y \in \Kd \\
    \Fb\{{\xb}\}, & \mbox{otherwise,}
    \end{cases}
\end{equation}
where $\hat\yb$ denotes the motion-corrupted $k$-space with the indices along the frequency encoding direction $k_x$ and phase encoding direction $k_y$, and $\xb$ is the motion-free image, $\Fb$ denotes the Fourier transform, $\Rb_\alpha$ denotes the rotation operation with the angle $\alpha$, $\Phi$ is the displacement in radian, and $\Kd$ is the phase encoding indices where the rotation or translation occurred.
 
Based on this assumption, the network is trained to reconstruct fully sampled motion-free images from randomly sub-sampled images along the phase encoding direction in which the corrupted $k$-space data can be removed in a probabilistic manner.
%During the inference step, the $k$-space samples of motion-corrupted images are randomly subsampled with several downsampling masks, so some sparse outliers are removed probabilistically.
%Then, downsampled images are reconstructed by the trained network and aggregated by averaging to acquire the final reconstructed image with reduced motion artifacts.
Although this method does not require simulated motion artifact images and shows improved performance, it has a limitation in that it is difficult to apply when the motion artifacts cannot be considered as sparse outliers in $k$-space.
Furthermore, because the index of outliers is not known, some outliers that are not removed by subsampling can remain in the reconstructed image.

\subsection{Proposed Method}
%The score-based diffusion model shows the impressive performance when the forward model is known in inverse problems.
%Moreover, it has been shown that the score-based diffusion model has strengths in high-frequency detail restoration in inverse problems \cite{song2021solving,chung2022come,chung2022score,saharia2022image}.
%However, in the case of MR motion artifact reduction, it is hard to identify the accurate forward model in Eq. \eqref{eq:motion} because the phase encoding indices with motion are not known.

\begin{figure}[!t]
\centerline{\includegraphics[width=0.9\columnwidth]{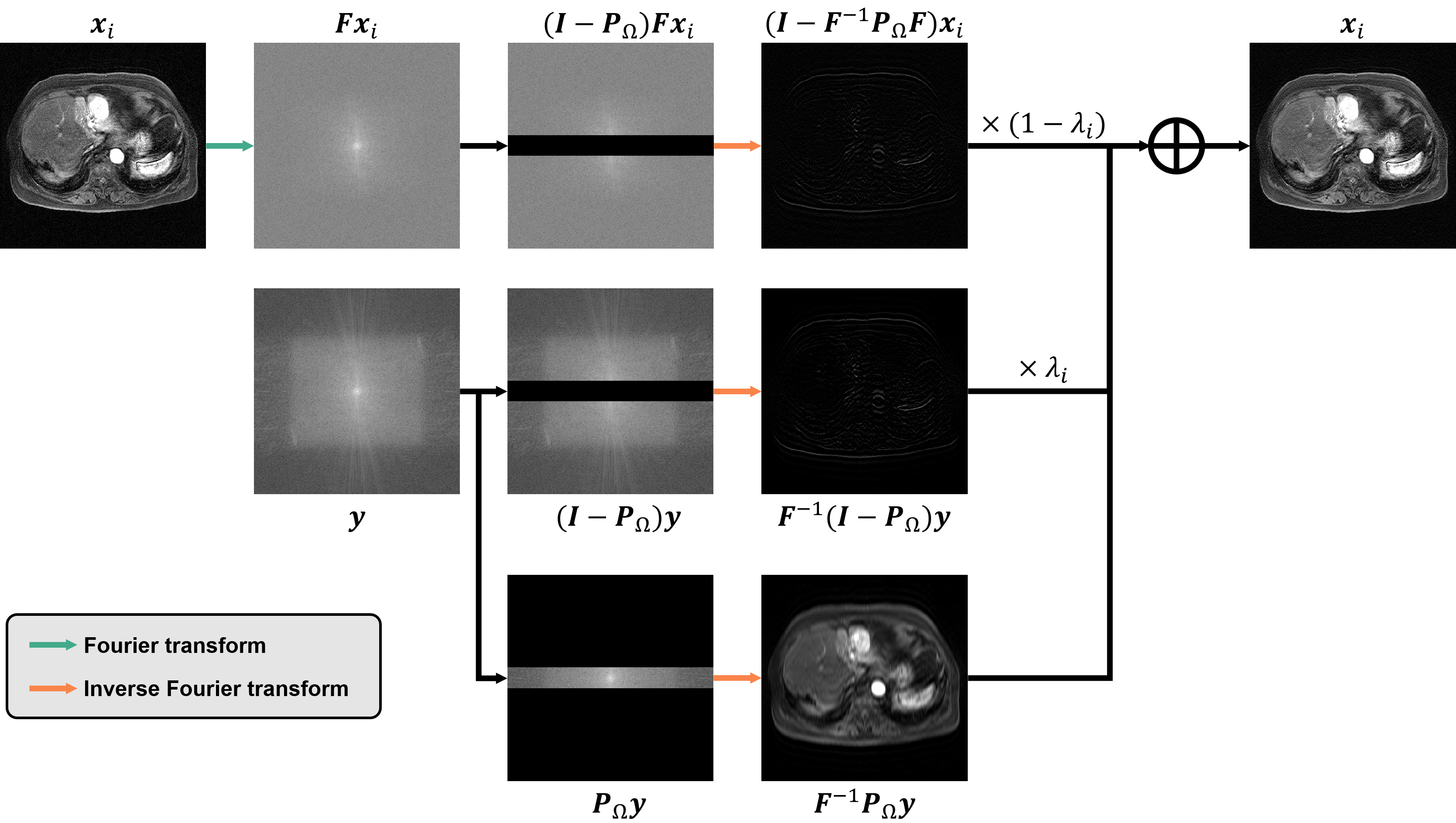}}
\caption{The data consistency step of the proposed method.}
\vspace{-0.5cm}
\label{fig:dc}
\end{figure}

Rather than using the sparse outlier assumption in Eq. \eqref{eq:motion}, our method is based on a more relaxed assumption that the motion artifacts in MRI mainly occur in the high-frequency region of $k$-space.
This is because $k$-space acquisition is usually performed first in the center region and motion occurs after a certain period after the start of acquisition so that $k$-space samples with motion artifacts generally appear in high-frequency regions.
Therefore, the high-frequency region of $k$-space should be corrected to remove the motion artifact.

The application of CCDF in Algorithm \ref{al:CCDF} starts from the one-step forward diffusion from the initialization.
% artifact-corruptedimage, which still provides a good initialization for the final artifact-corrected image.
Then, a n\"{a}ive way of using data consistency for reverse diffusion would be to impose the low-frequency region consistency:% would be imposed as the data consistency step:
%
%Because diffusion models can provide reconstructed images with high-frequency details in inverse problems, it can be done by CCDF in Algorithm \ref{al:CCDF} with the following data consistency step:
\begin{equation}\label{eq:dc_low}
    \xb_{i - 1} = (\Ib - \Fb^{-1}\Pb_\Omega\Fb)\xb'_{i - 1} + \Fb^{-1}\Pb_\Omega\yb,
\end{equation}
where $\Pb_\Omega$ is the operator that samples only the low-frequency region of $k$-space.
In other words, during reverse diffusion, the low-frequency region is maintained so that only the high-frequency region is corrected by the diffusion model.

However, because the high-frequency region of $k$-space also contains the information of details, the detailed structures of images can be altered or vanished if the data consistency step in Eq. \eqref{eq:dc_low} is applied directly.
To address this issue, we propose an annealed data consistency step to maintain high-frequency details of measurements as illustrated in Fig. \ref{fig:dc}:
\begin{equation}\label{eq:dc}
    \begin{split}
        \xb_{i - 1} &= (1 - \lambda_i)(\Ib - \Fb^{-1}\Pb_\Omega\Fb)\xb'_{i - 1}\\
        &+ \lambda_i \Fb^{-1}(\Ib - \Pb_\Omega)\yb + \Fb^{-1}\Pb_\Omega\yb,
    \end{split}
\end{equation}
where $\lambda_i \in [0, 1]$ is the annealing hyperparameter to control the weight of high-frequency components of the measurement.
Furthermore, as shown in Algorithm \ref{al:motion}, we choose relatively small $N'$, and repeat forward and reverse processes $M$ times so that the high-frequency components of the measurement are gradually added at each data consistency step.
%The effect of the choice of $N'$ and $M$ will be shown in Section \ref{sec:discussion}.

Here, Eq. \eqref{eq:dc} can be written as
\begin{equation}\nonumber
    \xb_{i - 1} = \Tb(\xb_{i - 1}) := \Ab\xb'_{i - 1} + \bb,
\end{equation}
where
\begin{equation}\nonumber
    \begin{split}
        &\Ab = (1 - \lambda_i)(\Ib - \Fb^{-1}\Pb_\Omega\Fb),\\ &\bb = \lambda_i \Fb^{-1}(\Ib - \Pb_\Omega) + \Fb^{-1}\Pb_\Omega.
    \end{split}
\end{equation}
Since $\norm{\Ib - \Fb^{-1}\Pb_\Omega\Fb} \leq 1$  \cite{chung2022come}, it is also true that $\norm{\Ab} = \norm{(1 - \lambda_i)(\Ib - \Fb^{-1}\Pb_\Omega\Fb)} \leq 1$.
Therefore, $\Tb$ is a non-expansive mapping, so it can accelerate the reverse diffusion process through the CCDF principle \cite{chung2022come}.
%The visualization of the data consistency step is .

%
%Therefore, our method does not require simulated or in vivo motion-corrupted data during the training.
%Our method is similar to Algorithm \ref{al:CCDF} with the data consistency step in Eq. \eqref{eq:dc}.
%The main difference is that the forward and reverse diffusion processes are repeated several times.
%One possible option is to choose large $N'$ and perform forward and reverse processes only once.
%However, we found that it is not suitable for the motion artifact reduction problem in our experiments.

% \setlength{\textfloatsep}{1pt}
\begin{algorithm}[!t]
\caption{MR Motion Artifact Reduction}\label{al:motion}
\begin{algorithmic}[1]
\Require{$\xb_0$, $\yb$, $N'$, $\{\sigma_i\}^{N'}_{i=1}$, $\{\epsilon_i\}^{N'}_{i=1}$, $\{\lambda_i\}^{N'}_{i=1}$, $\blmath{s_\theta}$}
\For{$j = 1$ to $M$}
\State{$\zb \sim \Nc(\blmath{0}, \Ib)$}
\State{$\xb_{N'} \leftarrow \xb_0 + \sigma_{N'}\zb$} \Comment{Forward diffusion}
\For{$i = N'$ to 1}
\State{$\xb'_{i - 1} \leftarrow \xb_i + (\sigma^2_i - \sigma^2_{i - 1})\blmath{s_\theta}(\xb_i, \sigma_i)$}
\State{$\zb \sim \Nc(\blmath{0}, \Ib)$}
\State{$\xb_{i - 1} \leftarrow \x'_{i - 1} + \sqrt{\sigma^2_i - \sigma^2_{i - 1}}\zb$} \Comment{Predictor}
\State{$\xb_{i - 1} \leftarrow (1 - \lambda_i)(\Ib - \Fb^{-1}\Pb_\Omega\Fb)\xb_{i - 1}$ \\
$\qquad\qquad\qquad + \lambda_i \Fb^{-1}(\Ib - \Pb_\Omega)\yb + \Fb^{-1}\Pb_\Omega\yb$\\\qquad} \Comment{Data consistency}
\State{$\zb \sim \Nc(\blmath{0}, \Ib)$}
\State{$\xb_{i - 1} \leftarrow \xb_{i - 1} + \epsilon_i\blmath{s_\theta}(\xb_i, \sigma_i) + \sqrt{2\epsilon_i}\zb$} \Comment{Corrector}
\State{$\xb_{i - 1} \leftarrow (1 - \lambda_i)(\Ib - \Fb^{-1}\Pb_\Omega\Fb)\xb_{i - 1}$ \\
$\qquad\qquad\qquad + \lambda_i \Fb^{-1}(\Ib - \Pb_\Omega)\yb + \Fb^{-1}\Pb_\Omega\yb$\\\qquad} \Comment{Data consistency}
\EndFor
\EndFor
\end{algorithmic}
\end{algorithm}

\subsection{Implementation Details}
In our implementation, we choose VE-SDE, which results in the following one-step forward sampling:
\begin{align}
\xb(t)=\xb(0)+\sigma(t) \zb
\end{align}
where $\zb\sim \Nc(0,\Ib)$ and $\xb(0)$ is the clean training data.
By plugging this in Eq.~\eqref{eq:objective}, we have the following cost function \cite{song2020score}:
\begin{equation}
    \begin{split}
        \min_\theta\Ed_t\Ed_{\xb(0)}&\Ed_{\xb(t)|\xb(0)}\Biggr[\\
        &\norm{\sigma(t)\blmath{s_\theta}(\xb(t), t) - \frac{\xb(t) - \xb(0)}{\sigma(t)}}_2^2\Biggr].
    \end{split}
\end{equation}
%Here, $\blmath{s_\theta}$ is trained with the objective function in Eq. \eqref{eq:objective}, 
%where $\xb_0$ is sampled from the motion-free images.
Here, we choose the number of discretized steps $N=1000$, and $\sigma_{\text{min}}$ and $\sigma_{\text{max}}$ in Eq. \eqref{eq:sigma} are set to $\sigma_{\text{min}}=0.01$ and $\sigma_{\text{max}}=50$, respectively.
We train the score model for 1.3M iterations and follow \cite{chung2022score} for the setting of other hyperparameters such as optimization, batch size, learning rate, gradient clipping, or exponential moving average.

In addition, for $N'$, $M$ and $\lambda_i$ in Algorithm \ref{al:motion}, we choose $N'=10$, $M=3$, and
\begin{equation}
    \lambda_i = \frac{\lambda_{N'}}{N' - 1}(i - 1),
\end{equation}
where $\lambda_{N'}=0.01$.
In other words, $\lambda_i$ linearly decreases to 0 as $i$ goes to 1, so the weight of high-frequency components of the measurement decreases as reverse diffusion proceeds.

In CCDF \cite{chung2022come}, it was shown that a better initialization provides faster reverse sampling.
Accordingly, the neural network (NN) initialization could be utilized if available as it is better than the original artifact-corrupted images. %, where the corrupted image is initialized by a pre-trained neural network.
%For example, $\xb_0$ in Algorithm \ref{al:motion} can be an output of the network of other motion artifact reduction algorithms.
%NN initialization does not require a large amount of computational cost or time, and it can largely reduce the sampling time of diffusion models.
Accordingly, we also employed NN initialization with \cite{armanious2020unsupervised} for the brain dataset, and \cite{oh2021unpaired} for the liver dataset.

\section{Methods}\label{sec:methods}
\subsection{Experimental Data}
In our experiments, we use two MR datasets.
The first dataset is the human connectome project (HCP) dataset which is the public dataset that contains human brain MR images.
This dataset is acquired by Siemens 3T system with 3D spin echo sequence, and the scan parameters are as follows: TR = 3200 ms, TE = 565 ms, echo train duration = 1105, matrix size = 320$\times$320, voxel size = 0.7$\times$0.7$\times$0.7 mm$^3$, and phase encoding direction = anterior-posterior.
Because the HCP dataset does not contain motion-corrupted images, it is used for quantitative evaluation with motion artifact simulation.
The score model is trained with 3000 motion-free MR images from 150 subjects, and other 800 images from 40 subjects are used for testing.

The second dataset is collected from Chungnam National University Hospital (CNUH), and it includes Gd-EOB-DTPA-enhanced liver MR images.
It is obtained by a 3T Philips Achieva MR system with the following scan parameters: TR = 3.1 ms, TE = 1.5 ms, flip angle = 10$^\circ$, field of view = 256$\times$256 mm$^2$, slice thickness/intersection gap = 2/0 mm, acquisition matrix = 320$\times$192, and phase encoding direction = anterior-posterior.
Also, dynamic imaging including various phases was obtained, but only arterial phase images are used for experiments because TSM usually occurs during the arterial phase.
The liver dataset consists of two groups, motion-free images, and motion-corrupted images.
For the training of the score model, 3097 motion-free images from 18 subjects are used.
After training, 444 simulated motion-corrupted images from 5 subjects are selected for the quantitative evaluation, and 38 MR volumes with in vivo motion-corrupted images are used for qualitative and radiologist evaluations.

\subsection{Artifact Simulation}
%Because it is difficult to obtain in vivo motion-corrupted and free image pairs,
For the quantitative evaluation, we used simulated motion artifacts.
The simulation was performed similarly to prior works \cite{tamada2020motion,oh2021unpaired}.
The first type of motion artifact that we simulate is random translation and rotation.
We simulate the first type of motion artifact with the HCP brain dataset.
The motion artifact with random translation and rotation can be simulated by Eq. \eqref{eq:motion} with
\begin{equation}\label{eq:random}
    % \begin{split}
    %     \alpha &= 
    %     \begin{cases}
    %     \alpha_{k_y}, & |k_y| > k_0 \\
    %     0, & \mbox{otherwise,}
    %     \end{cases},\\
    %     \Phi &= 
    %     \begin{cases}
    %     k_y\Delta_{k_y} + k_x\Delta_{k_x}, & |k_y| > k_0 \\
    %     0, & \mbox{otherwise,}
    %     \end{cases},
    % \end{split}
    (\alpha, \Phi) =
    \begin{cases}
        (\alpha_{k_y}, k_y\Delta_{k_y} + k_x\Delta_{k_x}), & |k_y| > k_0 \\
        (0, 0), & \mbox{otherwise,}
    \end{cases}
\end{equation}
where $\alpha_{k_y}$ denotes the rotation angle, $\Delta_{k_y}$ and $\Delta_{k_x}$ denote the degree of motion along $x$ and $y$ direction, respectively, and $k_0$ is the delay time of the phase error due to the centric $k$-space filling.
In our simulation, $k_0$ is fixed to $\pi/10$, $\alpha_{k_y}$ is randomly sampled from $[-2^\circ, 2^\circ]$, $\Delta_{k_y}$ and $\Delta_{k_x}$ are sampled from $[-1 \text{cm}, 1 \text{cm}]$ and $[-0.5 \text{cm}, 0.5 \text{cm}]$, respectively, at each $k$-space line.

The second type of simulated motion is respiratory motion, which appears as a sinusoidal function in $k$-space \cite{tamada2020motion,oh2021unpaired}:
\begin{equation}\label{eq:periodic}
    % \alpha = 0, \quad \Phi = 
    %     \begin{cases}
    %     k_y\Delta_{k_y} \sin(mk_y + n), & |k_y| > k_0 \\
    %     0 & \mbox{otherwise,}
    %     \end{cases},
    (\alpha, \Phi) =
    \begin{cases}
        (0, k_y\Delta_{k_y} \sin(mk_y + n)), & |k_y| > k_0 \\
        (0, 0), & \mbox{otherwise,}
    \end{cases}
\end{equation}
where $\Delta_{k_y}$, $m$, and $n$ denote the amplitude, period, and phase shift of the sinusoidal function, respectively.
Because the respiratory motion appears in abdominal MR images, we simulate it with the liver MR image dataset.
Parameters for the simulation are sampled as follows: $k_0 \sim U[\pi/10, \pi/5]$, $\Delta_{k_y} \sim U[1 \text{cm}, 1.5 \text{cm}]$, $m \sim U[0.1, 5.0]$, and $n \sim U[0, \pi/4]$, where $U[a, b]$ denotes the uniform distribution with the interval $[a, b]$

\subsection{Comparison Methods}
We compared our method with three state-of-the-art methods to verify the performance of the method.
The first comparison method is MARC \cite{tamada2020motion}, a method for reducing liver MRI motion artifacts.
Because it is a supervised method, we train MARC models using simulated motion-corrupted images with Eqs. \eqref{eq:random} and \eqref{eq:periodic}.

The second comparison method is Cycle-MedGAN V2.0 \cite{armanious2020unsupervised}, an unpaired deep learning method based on CycleGAN \cite{zhu2017unpaired}.
Cycle-MedGAN V2.0 can be trained with both simulated or in vivo motion-corrupted data, but we train it with only simulated motion-corrupted data because the training of Cycle-MedGAN V2.0 was unstable when using in vivo data.

We also employed the bootstrap subsampling and aggregation method in \cite{oh2021unpaired} as a comparison method.
Because this method requires only motion-free images during training, simulated or in vivo motion-corrupted images were not used during training.

\begin{figure*}[!t]
\centerline{\includegraphics[width=0.95\textwidth]{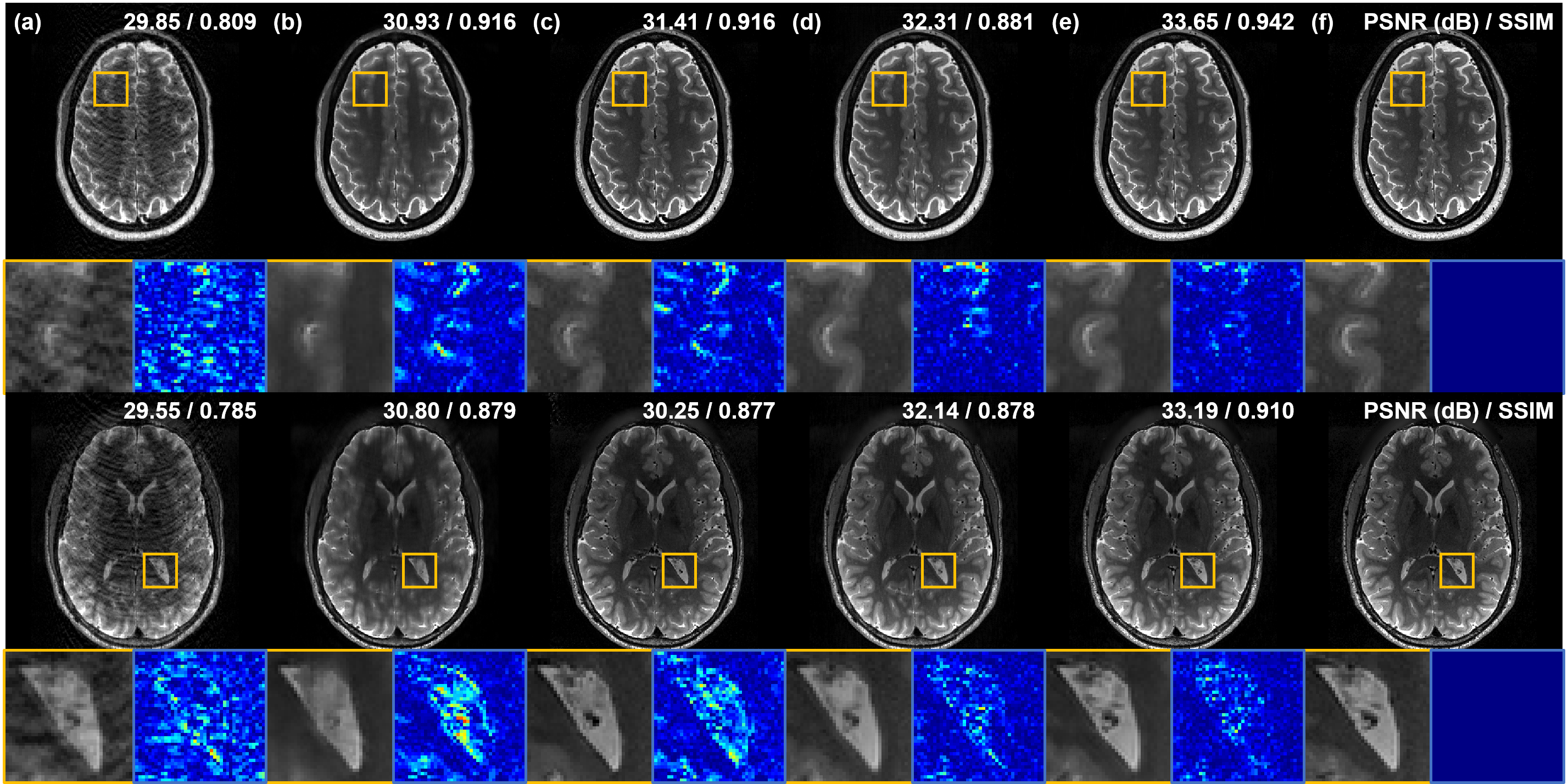}}
\caption{The simulated random motion artifact reduction results with the HCP brain dataset: (a) motion-corrupted input image, (b) MARC \cite{tamada2020motion}, (c) Cycle-MedGAN V2.0 \cite{armanious2020unsupervised}, (d) bootstrap subsampling and aggregation \cite{oh2021unpaired}, (e) the proposed method, and (f) motion-free label image.
The difference maps show the difference between each image and the label image.
PSNR and SSIM values of each image are shown in the corner of the images.}
%\vspace{-0.5cm}
\label{fig:results_brain_simul}
\end{figure*}

\subsection{Evaluation Methods}
For the quantitative evaluation, we used the peak signal-to-noise ratio (PSNR) and the structural similarity index metric (SSIM).
Because there is no ground truth matched with in vivo motion-corrupted images, the quantitative evaluation was performed with simulated motion-corrupted images.

In addition, we also conducted a clinical evaluation with the results using in vivo motion-corrupted data.
Specifically, a radiologist with 13 years of experience in abdominal MR imaging performed an analysis of the results of various methods.
The image analysis was conducted from various perspectives.
First, the performance in reducing motion artifacts is rated using a 5-point scoring system: 1 = non-diagnostic (severe artifacts causing impaired diagnostic capability of the readers); 2 = substantial artifacts with image quality decrease, but diagnostic performance impairment; 3 = mild artifacts, no significant (only mild) image quality disturbance; 4 = minimal artifacts, sharp image; 5 = no artifacts.
The image noise level is also evaluated with the following scoring system: 1 = non-diagnostic (severe noise causing impaired diagnostic capability of the readers); 2 = substantial noise with image quality decrease, but diagnostic performance impairment; 3 = mild noise, no significant (only mild) image quality disturbance; 4 = minimal noise; 5 = no noise.
Next, the blurring can be induced when reducing the motion artifact, so the rating of image blurring level is performed: 1 = non-diagnostic (severely pixelated texture causing impaired diagnostic capability of the readers); 2 = substantially pixelated, artificial sensation with concerns about the loss of normal texture, without diagnostic performance impairment; 3 = mildly pixelated, artificial sensation, without image quality decrease; 4 = minimal alteration of image texture; 5 = no alteration of image texture.
Furthermore, because the hepatic artery (HA) on the arterial phase should be visualized clearly, the vessel clarity is evaluated with a scoring system: 1 = not delineated due to motion or low signal-to-noise ratio (SNR); 2 = blur or decreased SNR; 3 = clear common hepatic artery (CHA) and proper hepatic artery (PHA), but blurred HA and gastroduodenal artery (GDA); 4 = entire HA is clearly visible, clear CHA, GDA, bilateral HA; 5 = strong contrast-to-noise ratio with score 4.
Last, the overall image quality is assessed by following scoring system: 1 = non-diagnostic; 2 = not satisfactory image quality, but re-examination is not needed; 3 = acceptable image quality (image quality may not be very good, but clinically acceptable); 4 = good image quality without significant artifact; 5 = excellent image quality without artifact and good spatial resolution.
The score is rated for each volume in all assessments.
Also, the results were presented to the radiologist in a random order without any labeling for a fair comparison.

\begin{figure*}[!t]
\centerline{\includegraphics[width=0.95\textwidth]{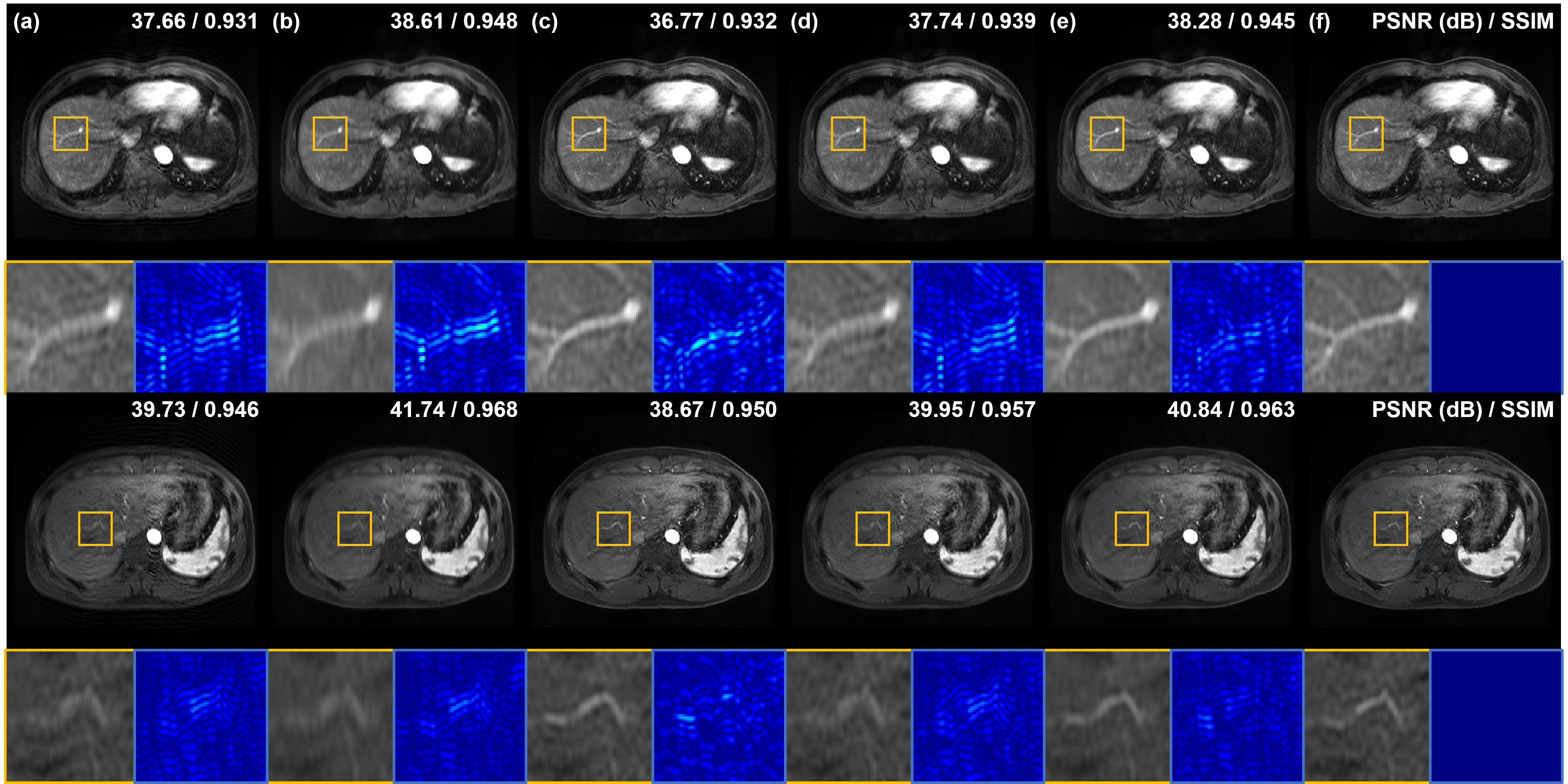}}
\caption{The simulated respiratory motion artifact reduction results with the CNUH liver dataset: (a) motion-corrupted input image, (b) MARC \cite{tamada2020motion}, (c) Cycle-MedGAN V2.0 \cite{armanious2020unsupervised}, (d) bootstrap subsampling and aggregation \cite{oh2021unpaired}, (e) the proposed method, and (f) motion-free label image.
The difference maps show the difference between each image and the label image.
PSNR and SSIM values of each image are shown in the corner of the images.}
%\vspace{-0.5cm}
\label{fig:results_liver_simul}
\end{figure*}

\section{Results}\label{sec:results}
\subsection{Results with Simulated Data}
Fig. \ref{fig:results_brain_simul} shows the motion artifact reduction results of various methods with random simulated motion-corrupted data.
As shown in Fig. \ref{fig:results_brain_simul}(a), it is hard to recognize detailed structures of brains due to motion artifacts.
MARC \cite{tamada2020motion} reduces the motion artifact but the output images of MARC are too blurry or smoothed (Fig. \ref{fig:results_brain_simul}(b)).
In the results of MARC, the boundary between gray matter and white matter is not clear (the first row in Fig. \ref{fig:results_brain_simul}(b)), and the structure of the choroid plexus is not properly restored (the second row in Fig. \ref{fig:results_brain_simul}(b)).
Next, in Fig. \ref{fig:results_brain_simul}(c) and (d), Cycle-MedGAN V2.0 \cite{armanious2020unsupervised} and bootstrap subsampling and aggregation \cite{oh2021unpaired} remove random motion artifacts significantly and show increased quantitative results compared to input images.
However, there are some differences between label images and outputs of Cycle-MedGAN V2.0 as shown in difference maps, and bootstrap subsampling and aggregation \cite{oh2021unpaired} shows blurrier edge details compared to label images.
On the other hand, as shown in Fig. \ref{fig:results_brain_simul}(e), the proposed method shows the best qualitative and quantitative results among all methods.
Especially, the proposed method shows the sharpest boundary between gray and white matters among methods as shown in the first row of Fig. \ref{fig:results_brain_simul}.

Next, we compare motion artifact reduction methods using simulated respiratory motion-corruption data.
In Fig. \ref{fig:results_liver_simul}(a), the vasculature of the liver is damaged or blurred due to motion artifacts.
Especially, artifacts appear most severe around blood vessels.
MARC removes motion artifacts and achieves high quantitative metric values, but the blood vessels still look blurry as shown in Fig. \ref{fig:results_liver_simul}(b).
On the other hand, Cycle-MedGAN V2.0 \cite{armanious2020unsupervised} sharp reconstructed results but the PSNR of results of Cycle-MedGAN V2.0 is lower than that of input images (\ref{fig:results_liver_simul}(c)).
It is maybe because Cycle-MedGAN V2.0 changes image intensity or details.
Results of bootstrap subsampling and aggregation \cite{oh2021unpaired} are shown in Fig. \ref{fig:results_liver_simul}(d), resulting in images with reduced motion artifacts and improved quantitative metrics compared to input images.
However, some motion artifacts near the blood vessels remain (the first row of Fig. \ref{fig:results_liver_simul}(d)), and it is hard to recognize the vessel due to blurring and remaining artifacts (the second row of Fig. \ref{fig:results_liver_simul}(d)).
Meanwhile, the proposed method shows the most similar restoration results to the label images as shown in Fig. \ref{fig:results_liver_simul}(e) and (f).
Specifically, the vascular structure is most clearly and accurately restored by the proposed method.
Furthermore, our method significantly reduces motion artifacts around the blood vessels compared to other methods.

TABLE \ref{tab:quantitative_metrics} shows the quantitative metric values of motion artifact reduction methods.
In experiments using simulated random motion-corrupted data, the proposed method achieves the highest PSNR and SSIM, and it is consistent with the qualitative results in Figs. \ref{fig:results_brain_simul} and \ref{fig:results_liver_simul}.
On the other hand, MARC shows the highest quantitative results when using simulated respiratory motion-corrupted data.
However, as confirmed in Figs. \ref{fig:results_brain_simul} and \ref{fig:results_liver_simul}, reconstructed images by MARC are extremely blurred, so the detailed structures are indistinguishable.
Compared to MARC, the proposed method removes the motion artifacts without losing information on image details.
Furthermore, the quantitative metric value of our method is the highest among that of unpaired/unsupervised methods.

\begin{table}[!h]
    \centering
    \caption{Quantitative results of various methods with simulated motion-corrupted data (Cycle: Cycle-MedGAN V2.0, BSA: Bootstrap Subsampling and Aggregation).}
    \resizebox{0.95\columnwidth}{!}{
    \begin{tabular}{c | c | c  c}
    \hline
    \   & Method                                        & PSNR (dB)            & SSIM               \\
    \hline \hline
    \multirow{5}{*}{\makecell{Brain\\Random motion}}
    & Input                                             & 27.83                & 0.751              \\
    & MARC \cite{tamada2020motion}                      & 29.29                & 0.891              \\
    & Cycle \cite{armanious2020unsupervised}            & 28.79                & \underline{0.894}  \\
    & BSA \cite{oh2021unpaired}                         & \underline{30.18}    & 0.839              \\
    & Proposed                                          & \textbf{31.40}       & \textbf{0.916}     \\
    \hline
    \multirow{5}{*}{\makecell{Liver\\Respiratory motion}}
    & Input                                             & 36.15                & 0.912              \\
    & MARC \cite{tamada2020motion}                      & \textbf{37.87}       & \textbf{0.947}     \\
    & Cycle \cite{armanious2020unsupervised}            & 35.54                & 0.926              \\
    & BSA \cite{oh2021unpaired}                         & 36.45                & 0.932              \\
    & Proposed                                          & \underline{37.01}    & \underline{0.940}  \\
    \hline
    \end{tabular}
    }
%    \vspace{-0.5cm}
    \label{tab:quantitative_metrics}
\end{table}

\begin{figure}[!t]
\centerline{\includegraphics[width=0.9\columnwidth]{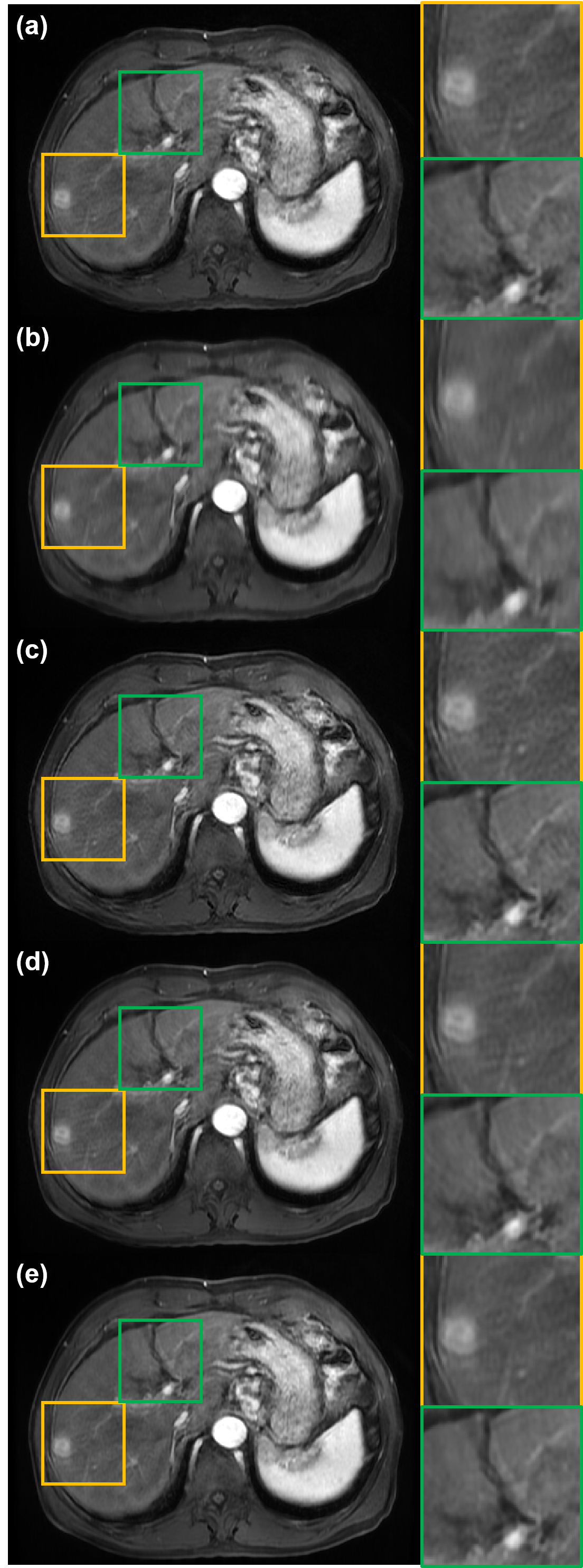}}
\caption{The in vivo motion artifact reduction results with the CNUH liver dataset: (a) motion-corrupted input image, (b) MARC \cite{tamada2020motion}, (c) Cycle-MedGAN V2.0 \cite{armanious2020unsupervised}, (d) bootstrap subsampling and aggregation \cite{oh2021unpaired}, and (e) the proposed method.}
%\vspace{-0.5cm}
\label{fig:results_liver_real}
\end{figure}

\subsection{Results with In Vivo Data}
Because the simulated motion artifacts only consider rigid motion artifacts, it should be verified that the method can also be applied to non-rigid in vivo motion artifact removal.
In Fig. \ref{fig:results_liver_real}(a), motion artifacts due to transient dyspnea degrade the quality of liver MR image.
We attempt to remove motion artifacts in Fig. \ref{fig:results_liver_real}(a), and results are shown in Fig. \ref{fig:results_liver_real}(b) to (e).
Again, MARC removes not only motion artifacts but also detailed structures of blood vessels, so the reconstructed image is extremely blurry (Fig. \ref{fig:results_liver_real}(b)).
Conversely, in Fig. \ref{fig:results_liver_real}(c), Cycle-MedGAN V2.0 makes the image sharper, but it also amplifies motion artifacts or noise in the input image.
Next, the bootstrap subsampling and aggregation method also fails to remove the motion artifacts.
Specifically, as shown in the yellow and green boxes of Fig. \ref{fig:results_liver_real}(d), motion artifacts around the blood vessels remain in the output image.
Unlike comparison methods, the proposed method successfully removes the motion artifacts and reduces the noise level of the input image.
Furthermore, our method reconstructs detailed structures.
For example, in the yellow box of Fig. \ref{fig:results_liver_real}(e), the sharpness of the lesion increased as the motion artifact disappeared.
Also, the vascular structure is recovered due to the reduction of motion artifacts as shown in the green box of Fig. \ref{fig:results_liver_real}(e).
Through the experiment using in vivo motion-corrupted data, we confirmed that the proposed method also removes in vivo motion artifacts that contain the non-rigid motion of patients.

\begin{table}[!h]
    \centering
    \caption{Clinical evaluation results of various methods with in vivo motion-corrupted data (average $\pm$ standard deviation) (Cycle: Cycle-MedGAN V2.0, BSA: Bootstrap Subsampling and Aggregation).
    Higher scores indicate higher performance.}
    \resizebox{\columnwidth}{!}{
    \begin{tabular}{c | c  c  c  c  c}
    \hline
    \ Method                                       & Motion artifact                & Noise                         & Blurring                        & Vessel clarity                  & Overall quality                 \\
    \hline \hline
    \ Input                                        & 3.03 $\pm$ 0.91                & 3.03 $\pm$ 0.68               & \underline{3.92} $\pm$ 0.71     & 3.45 $\pm$ 1.20                 & 3.00 $\pm$ 1.09                 \\
    \ MARC \cite{tamada2020motion}                 & 3.37 $\pm$ 0.79                & 3.34 $\pm$ 0.81               & 2.29 $\pm$ 0.87                 & 2.97 $\pm$ 1.15                 & 2.50 $\pm$ 1.03                 \\
    \ Cycle \cite{armanious2020unsupervised}       & 3.42 $\pm$ 0.92                & 3.13 $\pm$ 0.81               & \textbf{3.97} $\pm$ 0.94        & \underline{3.47} $\pm$ 1.18     & 3.21 $\pm$ 1.09                 \\
    \ BSA \cite{oh2021unpaired}                    & \underline{3.45} $\pm$ 1.22    & \underline{3.39} $\pm$ 0.75   & 3.89 $\pm$ 1.06                 & 3.45 $\pm$ 1.29                 & \underline{3.29} $\pm$ 1.18     \\
    \ Proposed                                     & \textbf{3.63} $\pm$ 1.10       & \textbf{3.58} $\pm$ 0.76      & \textbf{3.97} $\pm$ 0.91        & \textbf{3.71} $\pm$ 1.31        & \textbf{3.45} $\pm$ 1.25        \\
    \hline
    \end{tabular}
    }
%    \vspace{-0.5cm}
    \label{tab:clinical_evaluation}
\end{table}

\subsection{Clinical Evaluation}
Because it is impossible to quantitatively evaluate results using in vivo motion-corrupted datasets due to the lack of paired motion-free data, we evaluate motion artifact reduction results by clinical evaluation.

TABLE \ref{tab:clinical_evaluation} shows the scores by evaluating each method on various criteria.
MARC achieved scores of 3.37 and 3.34 in terms of motion artifact and noise evaluation, respectively, while input images score 3.03 in both evaluations.
These results indicate that MARC was good in motion artifact improvement or noise reduction.
However, MARC scored 2.29 in the blurring evaluation, which is lower than the score of input images (score: 3.92).
The blurring effect of MARC also can be confirmed in Fig. \ref{fig:results_liver_real}(b).
Therefore, the overall quality score of MARC (score: 2.50) is lower than that of input images (score: 3.00).
On the other hand, Cycle-MedGAN V2.0 got the highest score in the blurring evaluation (score: 3.97).
However, Cycle-MedGAN V2.0 scored 3.13 in noise evaluation, which is lower than the scores of other methods.
This high level of noise affects the image quality drop of Cycle-MedGAN V2.0, so Cycle-MedGAN V2.0 gets only 3.29 points in terms of the overall image quality.
As shown in Fig. \ref{fig:results_liver_real} and TABLE \ref{tab:clinical_evaluation}, the bootstrap subsampling and aggregation method shows higher scores than the other existing methods in most assessments.
However, the outputs of the bootstrap subsampling and aggregation method were slightly blurred, so its score was lower than the input images in the blurring evaluation.

While the other methods each showed drawbacks, the proposed method achieved the highest performance in all evaluations.
First, in terms of motion artifact removal, the proposed method achieves the highest score (score: 3.63) while other methods get similar lower scores (score: 3.37-3.45).
Next, our method scored 3.58 and 3.97 in the noise and blurring evaluations, respectively.
From these results, we confirm that our method does not amplify image noise level or blur output images through the clinical evaluation.
Moreover, the motion-corrupted input images scored 3.45 in terms of vessel clarity.
The proposed method shows a significant improvement in vessel clarity score (score: 3.71) while the vessel clarity of the other three methods is similar to or lower than that of motion-corrupted input images (score: 2.97-3.47).
Finally, our method gets the best score (score: 3.45) for overall image quality.
To sum up, the proposed method achieves the highest score in all clinical evaluations, and this result indicates that our method is useful in clinical practice.

% \begin{figure}[!t]
% \centerline{\includegraphics[width=0.65\columnwidth]{fig/ablation_lambda.png}}
% \caption{Ablation studies on the hyperparameter $\lambda_{N'}$ with in vivo motion-corrupted data: (a) motion-corrupted input image, (b) $\lambda_{N'} = 0$, (c) $\lambda_{N'} = 0.01$, and (d) $\lambda_{N'} = 0.1$.}
% \label{fig:ablation_lambda}
% \end{figure}

\section{Discussion}\label{sec:discussion}
\subsection{Comparison with Other Methods}
In Section \ref{sec:results}, it was verified that MARC \cite{tamada2020motion} generates blurry outputs in both simulation and in vivo study.
The blurring results may be a limitation of methods based on supervised learning.
Because the supervised learning minimizes the loss (e.g. L1, mean squared error (MSE)) between output and label, it achieves high quantitative results as shown in TABLE \ref{tab:quantitative_metrics}.
However, it can also lead to the loss of information on image details because L1 or MSE losses do not assure the perceptual quality of output images.

Unlike MARC, Cycle-MedGAN V2.0 \cite{armanious2020unsupervised} is an unpaired method that does not require paired input and label images.
Instead of using losses between input and label, it translates an image from one domain to another domain by utilizing cycle consistency loss and adversarial loss.
Because the discriminators of Cycle-MedGAN V2.0 distinguish real and fake generated images, the generators of Cycle-MedGAN V2.0 provide realistic images with sharp details.
However, we have confirmed that Cycle-MedGAN V2.0 also magnifies the artifacts or noise of images.
We conjecture that it is because the networks of Cycle-MedGAN V2.0 consider resolution degradation due to the motion artifacts to be the main difference between the two image domains.
Therefore, the networks of Cycle-MedGAN V2.0 try to improve resolution rather than eliminate motion artifacts.

Compared to the previous two methods, bootstrap subsampling and aggregation \cite{oh2021unpaired} showed stable qualitative and quantitative results.
Nevertheless, because \cite{oh2021unpaired} works under the assumption that the motion artifact appears as sparse outliers in $k$-space, the performance of this method is degraded if the assumption is not satisfied.
For example, we simulated the respiratory motion with Eq. \eqref{eq:periodic}, so the respiratory motion appears as a continuous sinusoidal form in $k$-space.
Because the motion did not appear as sparse outliers, the performance of \cite{oh2021unpaired} was dropped compared to when it works with simulated random motion-corrupted data.

On the other hand, our proposed method presented outstanding results compared to other comparison methods.
The proposed method successfully removes motion artifacts and retrieves high-frequency image details in both simulation and in vivo studies.

Nevertheless, our method is not free of limitations.
Because the score-based diffusion models require several steps of reverse diffusion, it takes a long time to generate outputs.
Although we utilized the CCDF algorithm to reduce the inference time, our method also requires several seconds as shown in TABLE \ref{tab:ablation}.
Therefore, the acceleration of the proposed method should be done for clinical use.

% \begin{figure}[!t]
% \centerline{\includegraphics[width=0.65\columnwidth]{fig/ablation_N.png}}
% \caption{Ablation studies on the hyperparameter $N'$ with in vivo motion-corrupted data: (a) motion-corrupted input image, (b) $N' = 1$, (c) $N' = 10$, and (d) $N' = 100$.}
% \label{fig:ablation_N}
% \end{figure}

% \begin{figure}[!t]
% \centerline{\includegraphics[width=0.65\columnwidth]{fig/ablation_M.png}}
% \caption{Ablation studies on the hyperparameter $M$ with in vivo motion-corrupted data: (a) motion-corrupted input image, (b) $M = 1$, (c) $M = 3$, and (d) $M = 5$.}
% \label{fig:ablation_M}
% \end{figure}

\begin{table}[!t]
    \centering
    \caption{Ablation studies on hyperparameters with simulated respiratory motion-corrupted data.
    The gray rows indicate the hyperparameters that are selected in our experiments.}
    \resizebox{0.95\columnwidth}{!}{
    \begin{tabular}{c | c | c  c  c}
    \hline
    \multicolumn{2}{c|}{Hyperparameters}     & PSNR (dB)                            & SSIM                                  & Time/image (sec)                    \\
    \hline \hline
    \multirow{3}{*}{$\lambda_{N'}$}
    & 0                                      & 36.36                                & 0.935                                 & 19.30                         \\
    & \cellcolor{Gray}0.01                   & \cellcolor{Gray}\textbf{37.01}       & \cellcolor{Gray}\textbf{0.940}        & \cellcolor{Gray}19.30         \\
    & 0.1                                    & 36.58                                & 0.927                                 & 19.30                         \\
    \hline
    \multirow{3}{*}{$N'$}
    & 1                                      & 36.45                                & 0.935                                 & 1.834                         \\
    & \cellcolor{Gray}10                     & \cellcolor{Gray}\textbf{37.01}       & \cellcolor{Gray}\textbf{0.940}        & \cellcolor{Gray}19.30         \\
    & 100                                    & 36.88                                & 0.938                                 & 195.6                         \\
    \hline
    \multirow{3}{*}{$M$}
    & 1                                      & 36.43                                & 0.934                                 & 6.358                         \\
    & \cellcolor{Gray}3                      & \cellcolor{Gray}37.01                & \cellcolor{Gray}0.940                 & \cellcolor{Gray}19.30         \\
    & 5                                      & \textbf{37.28}                       & \textbf{0.942}                        & 32.48                         \\
    \hline
    \multirow{2}{*}{$N' \times M$}
    % & 5 $\times$ 6                           & \textbf{37.12}                       & \textbf{0.942}                        & 19.30                         \\
    & \cellcolor{Gray}10 $\times$ 3          & \cellcolor{Gray}\textbf{37.01}       & \cellcolor{Gray}\textbf{0.940}        & \cellcolor{Gray}19.30         \\
    & 30 $\times$ 1                          & 36.90                                & 0.938                                 & 19.30                         \\
    \hline
    \end{tabular}
    }
%    \vspace{-0.5cm}
    \label{tab:ablation}
\end{table}

\subsection{Effects of  Annealing Hyperparameters}
In our method, we injected high-frequency components of measurements ($k$-space of motion-corrupted images) with the hyperparameter $\lambda_{N'}$ to preserve detailed structures of MR images.
To confirm the effect of high-frequency component injection, we conduct our method for simulated liver motion-corrupted images with various $\lambda_{N'}$.
As shown in TABLE \ref{tab:ablation}, the proposed method with $\lambda_{N'}=0$ shows lower quantitative results than the proposed method with $\lambda_{N'}=0.01$.
It is because detailed structures such as vessels cannot be reconstructed perfectly without high-frequency component injection.
When $\lambda_{N'} = 0.1$, the quantitative results drop again compared to results with $\lambda_{N'} = 0.01$.
We conjecture that it is because the high-frequency component of measurements also contains motion artifacts, and the remaining artifacts degrade the quality of reconstructed images.
Therefore, we choose to inject high-frequency components with $\lambda_{N'} = 0.01$ in our experiments.

% \begin{figure}[!t]
% \centerline{\includegraphics[width=0.65\columnwidth]{fig/ablation_NM.png}}
% \caption{Ablation studies on the combination of hyperparameters $N'$ and $M$ with in vivo motion-corrupted data: (a) motion-corrupted input image, (b) $N' = 5$, $M' = 6$, (c) $N' = 10$, $M' = 3$, and (d) $N' = 30$, $M' = 1$.}
% \label{fig:ablation_NM}
% \end{figure}

%\subsection{Effects of the Number of Steps of Reverse Diffusion}
Next, we also confirm the effect of the selection of $N'$.
When $N' = 1$, the motion artifacts remain in output images, so the quantitative results deteriorate.
On the other hand, our method also shows the degraded performance when $N' = 100$.
It may be because the structures that cannot be seen in the input image were generated during the iterations of the reverse diffusion process.
Moreover, the required inference time of the proposed method with $N' = 100$ is quite long as shown in TABLE \ref{tab:ablation}, so we choose $N' = 10$ that shows the best qualitative and quantitative performance.

%\subsection{Effects of the Number of Iterations}
Finally, the number of iterations of the reverse diffusion process $M$ is also one of the important hyperparameters of our method.
Through the experiments on $M$, we find that the proposed method cannot completely remove motion artifacts when $M = 1$.
On the other hand, when $M = 5$, the required inference time for one image is too long while the performance gain is negligible compared to when $M = 3$.
Therefore, $M = 3$ is selected in our experiments.

In addition, we also verify the effect of the combination of $N'$ and $M$.
The proposed method shows different results depending on the combination of $N'$ and $M$ as shown in TABLE \ref{tab:ablation}, even if it takes the same inference time.
The proposed method with $N' = 30$, $M' = 1$ shows lower quantitative performance compared to the method with $N' = 10$, $M' = 3$.
It is because the motion artifacts cannot be removed perfectly with only one iteration of the diffusion process even though $N'$ is large.
Through the experiment, we verify that the combination of $N' = 10$, $M' = 3$ is better than $N' = 30$, $M' = 1$ for the performance of our proposed method.

\section{Conclusion}\label{sec:conclusion}
In this paper, we proposed a novel MRI motion artifact reduction method using the annealed score-based diffusion model.
By applying the diffusion process iteratively and gradually imposing data consistency with high-frequency injection, the proposed method successfully reduced simulated and in vivo motion artifacts in MR images.
Furthermore, we verified that our method provides higher-quality images and more clinical meaning compared to other state-of-the-art deep learning methods.
We believe that our algorithm can be a useful framework for MRI motion artifact reduction.

\section{Acknowledgement}
This work was supported by Institute of Information \& communications Technology Planning \& Evaluation (IITP) grant funded by the Korea government (MSIT) (No.2019-0-00075, Artificial Intelligence Graduate School Program(KAIST)), National Research Foundation(NRF) of Korea grant NRF-2020R1A2B5B03001980, and by the KAIST Key Research Institute (Interdisciplinary Research Group) Project.

%\clearpage
\bibliographystyle{IEEEtran}
\bibliography{ref}

\end{document}

%% file: macros.tex
\usepackage{bm} % bold
\usepackage{xspace}
\newcommand{\xmath}[1] {\ensuremath{#1}\xspace}
\newcommand{\blmath}[1] {\xmath{\bm{#1}}}

\newcommand{\x}{\blmath{x}}

\newcommand{\norm}[1] {\xmath{\left\| #1 \right\|}}

\newcommand{\Ab}{{\blmath A}}

\newcommand{\Fb}{{\blmath F}}

\newcommand{\Ib}{{\blmath I}}

\newcommand{\Pb}{{\blmath P}}

\newcommand{\Rb}{{\blmath R}}

\newcommand{\Tb}{{\blmath T}}

\newcommand{\bb}{{\blmath b}}

\newcommand{\fb}{{\blmath f}}

\newcommand{\wb}{{\blmath w}}
\newcommand{\xb}{{\blmath x}}
\newcommand{\yb}{{\blmath y}}
\newcommand{\zb}{{\blmath z}}

\newcommand{\Nc}{\mathcal{N}}

\newcommand{\Rd}{{\mathbb R}}

\newcommand{\Ed}{{{\mathbb E}}}

\newcommand{\Kd}{\mathbb{K}}

\newcommand{\beq}{\begin{equation}}
\newcommand{\eeq}{\end{equation}}
\newcommand{\beqa}{\begin{eqnarray}}
\newcommand{\eeqa}{\end{eqnarray}}